# Quantum Technologies and Society: Towards a Different Spin

*Authors[1]:*

**Christopher Coenen**, Institute for Technology Assessment and Systems Analysis (ITAS) at Karlsruhe Institute of Technology, Germany, christopher.coenen@kit.edu

**Alexei Grinbaum**, CEA-Saclay/IRFU/Larsim, France, alexei.grinbaum@cea.fr

**Armin Grunwald**, Institute for Technology Assessment and Systems Analysis (ITAS) at Karlsruhe Institute of Technology, Germany, armin.grunwald@kit.edu

**Colin Milburn**, Departments of Science and Technology Studies, English, and Cinema and Digital Media, University of California, Davis, USA, cnmilburn@ucdavis.edu

**Pieter Vermaas**, Philosophy Department, Delft University of Technology, the Netherlands, p.e.vermaas@tudelft.nl (corresponding author)

*Abstract*:

Due primarily to technological advances over the last decade, quantum research has become a key priority area for science and technology policy all over the world. With this manifesto we wish to prevent quantum technology from running into fiascos of implementation at the interface of science and society. To this end, we identify key stumbling blocks and propose recommendations.

---

[1] All authors have with equal efforts contributed to this manifesto. The final writing was mainly done by Alexei Grinbaum, Colin Milburn and Pieter Vermaas. All authors commented on previous versions of the manuscript, and all authors read and approved the final manuscript.

It would seem that the second quantum revolution is well under way. Innovative quantum technologies are poised to introduce novel possibilities and challenges for society. New algorithms based on quantum superposition and entanglement, new protocols for secure communication, new sensors and highly precise devices based on quantum resources will bring a new wave of industrial applications. Some commentators have foreseen an imminent "gold rush," a global drive to extract wealth from the quantum realm.[2] Others have observed the onset of another high-tech "arms race," noting not merely the potential military uses of quantum technologies but a more pervasive sense of international competition and conflict that spurs innovators and investors to action.[3] Such claims, such tropes, are familiar elements from the repertoire of cultural narratives about new and emerging technologies: standard plotlines that not only describe but also prescribe the course of things to come. Perhaps, however, we might find other ways of orchestrating the new era of quantum technologies, rather than simply recreating the patterns and clichés of the past.

As researchers in the fields of science and technology studies, scholars of the history and philosophy of the physical sciences, we have extensive experience in studying emerging technologies as well as public debates about their implications. We are each currently involved in several different science education initiatives and policy discussions pertaining to new quantum technologies. Our aim with this manifesto is to enrich the quantum technology debate and prevent it from running into communicative dilemmas or fiascos of implementation at the interface of science and society. We advise critical reflection and care in articulating the terms, conceptual frames, and evolving trajectories of the debate, attentive to the performative and constructive force of inherited language and regurgitated cultural narratives, intentional or otherwise. This advisement holds for technical research reports as much as popular science journalism—and especially for developing policy agendas, pedagogical curricula, institutional roadmaps, and technical standards. Language matters, because the way we talk about things sets a horizon of expectation for how we go about doing things. And the way we do things matters, even more.

\*

Due primarily to technological advances over the last decade, quantum research has escalated to become a key priority area for science and technology policy all over the world, with significant amounts of taxpayer money—to say nothing of venture capital and Big Tech investments—flowing into the field. As with other new and emerging innovations that portend widespread social impacts, appealing to public as well as private interests and capturing a large slice of popular attention, it is imperative to examine any factors that might potentially impede or circumscribe the democratic potentials of this new quantum age.

Quantum technology is regularly draped in a rhetoric of the unprecedented—a sense of estrangement and radical novelty, which derives from the longer history of quantum physics and the spooky weirdness associated with quantum phenomena. Richard Feynman, who among other achievements was one of the founders of quantum computing [6], famously wrote, "I think I can safely say that nobody understands quantum mechanics." [7] This sentiment still has many aficionados. Quantum physics has often been accused of being counterintuitive and incomprehensible—attributes which physicists embrace to claim their authority on the quantum domain, which philosophers of science emphasize to explore how classical conceptions of reality are eroded, and which popular culture often refigures as esoteric access to the ineffable. Quantum physics is also frequently portrayed as the pinnacle of the natural sciences by virtue of its claim to describe the most fundamental level of nature, its formal articulations that are as elegant as they are opaque to those untrained in its mathematical techniques, and its first wave of applications in the nuclear industry and atomic weaponry, through which quantum theory became directly involved in the geopolitical ordering of the world in the aftermath of World War II. Accordingly, the presentation of quantum technology is now laden with the language of the extraordinary, its novelty overdetermined by these connotations and historical associations. For example, the physicist Ron Folman at Ben-Gurion University of the Negev has said, "Quantum theory includes a very strange set of rules and its quirkiness will result in revolutionary quantum technology with vastly different capabilities than what we are used to."[4] Similarly, Stephan Ritter and Jürgen Stuhler at Toptica Photonics AG have written, "The most fragile properties of quantum systems, often perceived as counterintuitive or even spooky, are actually the source of radically new technologies." [9] Quantum technologies are promoted as being socially and economically disruptive insofar as quantum phenomena appear counterintuitive and incomprehensible, accessing weird forces unavailable to classical technologies—surfing a wave of so-called quantum woo. These capabilities of quantum technology are increasingly framed as shaping the geopolitical balance of the twenty-first century—a new battle for scientific dominance, as suggested by the French President Emmanuel Macron in 2021:

> "We are aware of the Chinese and American competition but—given the disciplines that are key to the quantum, given what has already been done by our scientific community, given the involvement of our fundamental research and industrial stakeholders—today we have the recipes for success and for being among the top players in this battle."[5]

---

[2] For examples, see [1, 2].
[3] For examples, see [3, 4, 5].

Yet, even if quantum physics may be counterintuitive, there is no reason for quantum technologies to be promoted as inscrutable or enigmatic. Transistors, after all, are based on quantum phenomena, and ordinary computers that are built out of them are not perceived as towers of mystery. Similarly, communication devices or sensors based on quantum technologies are likely to present incremental improvements to already existing systems and practices. Of course, new technologies keep appearing in a constant stream of innovation, and the accompanying promises often seek to impress above and beyond the scope of actual achievements. The esoteric halo around quantum mechanics, although it has enjoyed a certain popularity, should not mean that the users of quantum technologies will have to live in a world of mystery and give up on any attempt to comprehend. On the contrary, we need responsible heuristic techniques to make quantum technologies into familiar elements of the citizen's technological palette [11]. Instead of trite appeals to Schrödinger's cat, we need to speak about quantum resources that enable new solutions to real-world problems. Instead of indulging in enigma by focusing on the meaning of the wave function, the measurement problem, or treading into other fundamental tenets of mathematical formalism, we should attend to useful resources such as entanglement and protocols such as quantum key distribution or quantum teleportation [12]. Rather than being perceived as uncanny guests from the quantum realm, these basic operations should become common cultural terms known to everyone.

---

[4] Ron Folman quoted in [8].
[5] Emmanuel Macron [10]; our translation. The original quotation in French: "Nous savons la compétition chinoise, américaine, mais compte tenu des disciplines qui sont clés dans le quantique, compte tenu de ce qui a déjà été fait par notre communauté scientifique, compte tenu de l'implication de nos acteurs de recherche fondamentale et des industriels, nous avons les recettes aujourd'hui pour réussir et être parmi les tout meilleurs dans cette bataille."

We would be loath to present quantum technologies in a sober scientific way at the cost of stripping quantum theory of all its wonderful properties. But those properties should not be detached from the tradition in which they were originally discussed by the wide-ranging thinkers of previous generations, including scientists such as Niels Bohr or Albert Einstein, philosophers such as Alfred North Whitehead or W. V. O. Quine, historians such as Max Jammer, psychologists such as C. G. Jung, and theologians such as John Polkinghorne. They each had a lot to say about the impact of quantum mechanics on our worldview, reflecting keen awareness of the cultural significance of science and sensitive appreciation for modern technology as an embedded human practice. Following suit, the broader significance of quantum theory needs to be transferred over to quantum information and quantum technologies. Once they become "normal" science—to use a famous expression by Thomas Kuhn—the aura of mystery will diminish, certainly, but the rich intellectual tradition must nevertheless be preserved and renewed. If this does not happen, the risk is that the cultural vacuum will be filled with nonsensical and possibly dangerous pseudoscientific ideas. The field of quantum technologies needs to be able to demonstrate not only the high-tech "perilous leaps," in the words of the poet W. H. Auden, but also cultural continuity.

Hence our stance: we are not simply advocating for improved modes of science communication and pedagogy—though, of course, innovative ways of disseminating knowledge about quantum phenomena and their technological applications are also necessary. We applaud the efforts of those in the field who have prioritized public science communication and educational initiatives.[6] However, we emphasize that, to whatever degree quantum technology research may rely on public funding and public good will, researchers and innovators must engage transparently with society to find out how they can resonate with public concerns and address social uncertainties. Quantum innovations developed in corporate environments, including the increasingly corporate zones of research universities, may predictably emerge under a shroud of mystery—one woven more from the warp of intellectual property regimes than from any inherent obscurity of the quantum realm. However, to the furthest extent possible, the open disclosure and active sharing of knowledge will be crucial for broad societal engagement with the implications of emerging quantum technologies, in ways that allow for responsive policy assessments and anticipatory governance [14, 15, 16. 17].

It is not helpful, after all, to proclaim the disruptive potential of quantum technologies based on promises alone. A lucid and modest approach that attends to constraints as much as opportunities is often better than selling unsubstantiated hype [18]. Overstretched promises may eventually come to haunt the very scientists who are making them. While there is an important and vital role for speculative visions in guiding research trajectories and opening inquiries into previously unforeseen possibilities, alluring promises cast in the language of profiteering ventures, especially those forecasting quick returns in the quantum gold rush, also carry a risk of backlash, deflated hopes, and broken development pathways. In certain cases, the financialization of forward-looking promises may even preempt innovation, triggering an internal seizure that forestalls technical breakthroughs before they have even been made [19, 20, 21, 22, 23, 24].

---

[6] For example, see [13].

If quantum technologies are to be meaningfully disruptive, they should not be developed in ways that merely extend the status quo, uncritically reproducing prevailing social conditions and unsustainable values into the quantum future. In the field of quantum computing, for example, the concept of "quantum supremacy"—designating a threshold event, a breakthrough achieved when a quantum computer efficiently solves a problem that cannot feasibly be solved by a classical computer—has recently come under scrutiny, due to the associations of the term "supremacy" with discourses of imperialism, colonialism, and, especially in anglophone contexts, the recalcitrant residues of white supremacy.[7] We may question, however, whether proposals to substitute alternative synonyms shorn of these odious connotations would actually represent a meaningful difference—at least, without also addressing the social and epistemic contexts in which such concepts operate. Replacing the terminology of "quantum supremacy" with "quantum advantage" or "quantum primacy," for example, would merely shift the semantic terrain from the domain of imperialism to the competitive markets of neoliberalism or the valoration of bare exceptionalism. In other words, the change in vocabulary terms would do little to address the underlying ideological values that make quantum supremacy such an appealing benchmark in the first place, as a measure of implicit competition—not only between quantum computing and classical computing, but more insidiously, between institutions, corporations, governments, worldviews. To be sure, quantum supremacy serves as a metonym for the anticipated economic and social advantages that will accrue to the first nation-states or economic zones to successfully develop, implement, and commercialize quantum technologies.

This much is suggested in several policy documents that have appeared in recent years. For instance, the 2016 *Quantum Manifesto*, [28] co-authored by representatives of the European Commission, the Dutch Ministry of Economic Affairs, Innovate UK, and three European universities, stresses the imagined superiority of quantum technologies: "Quantum computers are expected to be able to solve, in a few minutes, problems that are unsolvable by the supercomputers of today and tomorrow." The technological superiority of such inventions would seem to both reflect and reinforce the superiority of the people who develop them— hence, the necessity for "an ambitious European initiative in quantum technologies, needed to ensure Europe's leading role in a technological revolution now under way." If the difference that quantum makes—whether cast in the language of supremacy, advantage, or primacy— seems to entail "transformative applications" and "revolutionary new technologies" with resounding consequences, insisting that those who get there quickly will have some competitive advantage over those who are slower to catch up, then we are once again situated in an imaginary geopolitical race, a high-tech battle for dominance: "A world-wide race for technology and talent has started, as the strategic and economic stakes are high. Other parts of the world are speeding up and Europe cannot afford to lag behind ..." The implicit narrative here is less about competitive innovation as a means to global prosperity than about chauvinistic rivalry, a batrachomyomachia with grim prospects for the ongoing value of scientific cooperation.

---

[7] On the concept of quantum supremacy, see [25]. For a critique of the term's colonialist overtones, see [26]. For a critique of the term's semiotic echo of white supremacy, see [27].

Of course, we have been here before—but perhaps this time, let us resist the reinscription of competition into our technoscientific concepts and their implementation. Rather than yet another technological revolution figured in terms of combat and measures of superiority, the nascence of quantum technologies could instead afford other options—perhaps framed instead in terms of entanglement, contextuality, collaboration, and open science.[8]

*

As usual, it is easier said than done. But if we are to take a different spin this time around, if we aspire for quantum technologies to respect the values of transparency, fairness, and inclusivity, it will be necessary among other things to develop methods for speaking about quantum resources that do not simply rehearse familiar tropes or destine new technological developments to repeat the toxic patterns of history. Greater attentiveness to the sociotechnical specificities of quantum technologies will ultimately make them less mysterious and better understandable to the public—and also more resistant to established geopolitical scripts. We therefore conclude with some practical recommendations. Quantum technologies should be:

- Comprehensible: present quantum technologies in ways that are legible, honest, and publicly accountable, focusing on applications that may have particular advantages and disadvantages for different groups of stakeholders.
- Specific: instead of relying on generic metahistorical narratives and sociological clichés, attend to more specific innovation pathways and sociotechnical processes, denoting application fields, relevant actors, development strategies, legal and ethical considerations, and scientific challenges.
- Open: make quantum technologies available to communities beyond early adopter states, start-ups and Big Tech companies, and avoid tunnel visions that fixate on the first commercial use-cases for industry or applications that give the US, China or EU geopolitical advantages.
- Accessible: enhance the diversity of the field's workforce, design implementations that anticipate and support the greatest variety of users and contexts, and ensure that countries of the Global South have access to applications of quantum technology that are relevant for meeting specific needs, mindful of any disparities that may widen gaps or exacerbate inequalities.
- Responsible: involve sustainability research, technology assessment, and practices of responsible research and innovation to investigate possible long-term effects of quantum technologies, including unintended consequences, with an eye to the interests of future generations as much as our own.

---

[8] See, for example, [29, 30, 31, 32].

- Culturally embedded: develop outreach efforts and participatory opportunities for citizens that speak to implications of quantum technologies in the popular imagination, in different cultural contexts.
- Meaningful: engage with a greater variety of societal needs, hopes, and concerns, steering the development of quantum technologies toward applications that are meaningful not only for industry but also for society.

If quantum physics truly offers counterintuitive ways of apprehending the world, allowing us to perceive things differently, then the future of quantum technologies should not be approached as a zero-sum game of winners and losers but instead an opportunity for a different game entirely, reimagined in terms of nonbinary thinking and complementarity [33, 34, 35]. A better embrace between quantum technologies and society is both theoretically grounded and empirically feasible. The choices we make about these issues now, inevitably, will affect the future.